\documentclass[twocolumn,amsmath,amssymb]{revtex4}
\pdfoutput=1
\usepackage{latexsym}
\usepackage{amsmath}
\usepackage{soul,xcolor}
\usepackage{times}
\usepackage{amssymb}
\usepackage{fancyheadings}
\usepackage[T1]{fontenc}
\usepackage[utf8]{inputenc}
\usepackage{url}
\usepackage{hyperref}
\usepackage{color}
\usepackage{graphicx}
 \usepackage{epsfig}
\usepackage{mathptmx}
\usepackage{bm}
\usepackage{array}
\usepackage{algorithm}
\usepackage{algorithmic}
\usepackage{url}
\usepackage{hyperref}
\usepackage{color}
\usepackage{soul,xcolor}
\setstcolor{red}

\usepackage{lscape}
\usepackage{draftcopy}

\begin{document}
\title{Effect of hidden geometry  and  higher-order interactions on the synchronization and hysteresis behaviour  of phase oscillators on 5-cliques simplicial assemblies}
\author{Samir Sahoo$^{2,5}$, Bosiljka Tadi\'c$^{1,4}$, Malayaja Chutani$^3$ and Neelima Gupte$^{3,5}$}
\affiliation{$^1$Department of Theoretical Physics, Jo\v zef Stefan Institute,
Jamova 39, Ljubljana, Slovenia}
\affiliation{$^2$Department of Applied Mechanics, Indian Institute of Technology
  Madras, Chennai 600036, India}
\affiliation{$^3$Department of Physics, Indian Institute of Technology
  Madras, Chennai 600036, India}
\affiliation{$^4$Complexity Science Hub, Josephstaedterstrasse 39,
    Vienna, Austria}
\affiliation{$^5$Center for Complex Systems \& Dynamics, Indian Institute of Technology Madras, Chennai 600036, India}
\vspace*{3mm}
%%%%%%%%%%%%%%
\date{\today}

\begin{abstract}
\noindent 
The hidden geometry of simplicial complexes can influence the
collective dynamics of nodes in different ways depending on the
simplex-based interactions of various orders and competition between
local and global structural features. We study a system of phase
oscillators attached to nodes of 4-dimensional simplicial complexes and interacting via positive/negative edges-based pairwise  $K_1$ and triangle-based triple $K_2\geq 0$ couplings. Three prototypal simplicial complexes are grown by aggregation of 5-cliques, controlled by the chemical affinity parameter $\nu$, resulting in sparse, mixed, and compact architecture, all of which have  1-hyperbolic graphs but different spectral dimensions. By changing the interaction strength $K_1\in[-4,2]$ along the forward and backward sweeps, we numerically determine individual phases of each oscillator and a global order parameter to measure the level of synchronisation. 
Our results reveal how different architectures of simplicial
complexes, in conjunction with the interactions and internal-frequency
distributions, impact the shape of the hysteresis loop and lead to
patterns of locally synchronised groups that  hinder global
network synchronisation. 
Remarkably, these groups are differently affected by the size
of the shared faces between neighbouring 5-cliques and the presence of higher-order interactions. At $K_1<0$,  partial synchronisation is much higher in the compact community than in the assemblies of cliques sharing single nodes, at least occasionally. These structures also partially desynchronise at a lower triangle-based coupling $K_2$ than the compact assembly. Broadening of the internal frequency distribution gradually reduces the synchronisation level in the mixed and sparse communities, even at positive pairwise couplings. The order-parameter fluctuations in these partially synchronised states are quasi-cyclical with higher harmonics, described by multifractal analysis and broad singularity spectra.

\end{abstract}
\maketitle
\section{Introduction\label{sec:intro}}
Mapping complex systems onto networks that embody functional connections among the system's constitutive elements often involves higher-order couplings, which induces more complex geometries.
Identifying these hidden geometry features of various orders and
assessing their impact on the system's dynamics is currently in the
focus of network's theory \cite{HOC_review2023boccaletti,NetGeometry2021,HOC_PercGhos2022review} and its applications from the brain \cite{HC_cliquecavities_JCompNeurosci2018,we-Brain_SciRep2019,we-Brain_SciRep2020},
to large-scale social systems and  emergent networks
\cite{we_PhysA2015,we_Brain2016,we_Tags2016,HOC_evolution2021} to the   materials design \cite{AT_Materials_Jap2016,SC_we_SciRep2018,SC_we_PRE2020}.
These complex topologies can be described by algebraic topology
\cite{kozlov-book,Q-analysis-JJ} with identifying full graphs
(cliques) of all sizes that are present  as well as the faces that they
share with other cliques to make the actual simplicial complex. In
this context,  the underlying network (1-skeleton of the simplicial
complex) is made of the edges connecting two nodes, which may appear
as faces of the order one of larger simplexes (triangles, tetrahedrons, etc.). 

In the theory of complexity---the emergence of new features at a
larger scale, as a key property of a complex system, can be associated with
the collective dynamic fluctuations. Such dynamic phenomena,
through interactions among dynamical units,  appear with long-range spatiotemporal correlations that are characteristics of critical
states, either with a dynamical phase transition or as 
self-organised critical attractors in driven nonlinear dynamics; see a brief survey in
\cite{SOC_weDynamics2021} and references there. The underlying
simplicial structure can provide multiple interactions among dynamical
units associated with the network nodes.   Particularly the pairwise
interactions occur along the network's edges; meanwhile, the higher
interactions can be geometrically embedded into the triangles,
tetrahedrons and higher faces up to the largest simplex found in the
structure. The two leading (i.e., pairwise and
triangle-based) interactions are expected to impact the collective
dynamics critically, given the renormalisation-group theory \cite{RG_CNets2011frontiers,RG_spinglassParisi2010}; however, this
question remains open in finite systems and complex geometries.
The actual impact of these interactions also depends on the type of
the dynamics of interacting units. For example, the higher-order interactions, e.g.,
within large cliques  may enable the fast spreading of diseases
\cite{HOC_contagAA_PRX2020}, and enhance traveling waves in networks
of neurons but without pathological full synchronised states \cite{HOC_neuronsnets2019}.
On the other side, the triangle-based couplings induce geometric
frustration with long-range effects in spin kinetics \cite{spinfrustrations2015,SC_we_Entropy2020,SC_HOCweEPL2020}.

Phase synchronisation among many interacting units is a prototypal nonlinear  dynamics model to
study the cooperative phenomena in many complex systems, including
applications in neuroscience, engineering, etc.\
\cite{SYNC_Neuralmemory2011review,SYNCH_frustr_RoySoc2014,Rodrigues2016_sync_review}.
Current research on the influence of higher order coupling on
synchronization processes focus on: searching for conditions of perfect synchronization among oscillators on nodes of general networks in the presence of a $p$-point
interactions \cite{SC_HOCsynchroperfect_arxiv2023}; 
conditions for full synchronization between
topological signals associated with simplexes of different order; see recent work
\cite{SYNC_globalGinestraPRL23} and references there; and
understanding the nature of synchronisation--desynchronisation processes of the
oscillators at nodes of simplicial complexes with
geometric interactions embedded into simplexe's faces of different order
\cite{Arenas_NatureComm,HOC-Synch_Ginestra2020,SC_Sfboccaletti2021,SC_synchro_wePRE2021,SC_we_HCsynchro2022};
Our work belongs to the latter class of problems.
In this context, key questions concern the emergence and disappearance of collective dynamic behavior, measured by the order parameter on the hysteresis loop, when the strengths of the various interactions embedded in the geometry vary. 
For example, the presence of triangle-based interactions is understood
to disrupt the order promoted by increasing the strength of pairwise
couplings, and can cause sudden desynchronization \cite{Arenas_NatureComm,millan2019synchronization,SC_synchro_wePRE2021}.
Furthermore, the occurrence of \textit{partially synchronised} phases with negative pair interactions is another striking feature of geometric interactions on simplicial complexes; see, for example,
\cite{SC_synchro_wePRE2021} and references there. 
As a special case, the frustrated synchronisation
\cite{SYNC_FrustrExplosive2020} is often attributed to complex
structures, e.g., in brain networks, where higher geometries are expected
to play an important  role \cite{SC_we_HCsynchro2022,HOC_neuronsnets2019}.
Theoretically, the influence of topology on diffusive processes can be captured to a good extent by spectral analysis of the network
\cite{we-spectra2009,Spectra-topologyPRE2017,SC_we-PREspectra2019,Milan2018,millan2019synchronization}.
A suitable measure is the spectral dimension $d_s$ derived from the
eigenvalue spectrum of the Laplace operator associated with the
network adjacency matrix; higher-order combinatorial Laplacians
\cite{HOC_Laplacians_danijela2013} are adequate for the diffusion of
topological signals on simplicial complexes;  see \cite{SYNC_globalGinestraPRL23}.
In the context of phase synchronization, it has been understood that networks with $d_s\geq 4$ are required to enable stable global synchronization,
while such states cannot be reached when $d_s\lesssim 2$.
Unlike global synchronization, where conditions can be
formulated analytically, as in the above references\
\cite{SC_HOCsynchroperfect_arxiv2023,SYNC_globalGinestraPRL23}, the
origin of partial synchronization on simplicial complexes and the
nature of the underlying dynamical states are more elusive.
In addition to the topological dimension of a simplicial
complex, the role of its architecture in synchronization processes
remains unclear, especially in relation to by the presence of higher order interactions and the internal
inhomogeneity of the nodes' dynamics.

In this work, we tackle these questions by studying the synchronisation and desynchronisation processes on several 4-dimensional  simplicial complexes of a controlled architecture, all assembled by 5-cliques as building blocks
\cite{SC_we_SciRep2018}. Changing a controlled
parameter, as explained in detail in the following section, the
assemblies of 5-cliques with different architecture and spectral dimension
\cite{SC_we-PREspectra2019} are grown, which appear to have diverse impacts on phase synchronisation among the oscillators at their nodes. More specifically, we analyse the hysteresis loop by varying the pairwise
interaction from negative to positive values and back, with(out) the
three-phase couplings embedded into triangles of the actual complex. 
The network's ability to reach complete synchrony at the positive pairwise couplings and partial synchrony and incomplete desynchronisation by negative interactions depends on the simplicial architecture
corroborated with the distribution of internal frequencies. Remarkably, the level of partial synchronisation observed at negative pairwise couplings can be associated with the minimal size of the faces shared among
neighbouring 5-cliques and is virtually independent of the presence
of higher-order interactions and the actual frequency distribution.
The multifractal fluctuations of the order parameter are associated
with these partially synchronised states, emerging through roughly
synchronised small clusters. The organisation of clusters
is sensitive to the triangle-embedded interactions, even for the
global order parameter in the same range.

In section \ref{sec:nets}, we present  three considered simplicial
complexes and their structural features relevant to this work. 
Section \ref{sec:synchro} introduces the dynamical model with the leading pairwise and triangle-based
interactions on these complexes, in \ref{sec:simulations};
then in subsections \ref{sec:omega1}, \ref{sec:patterns} and
\ref{sec:gauss}, we give the results regarding the
hysteresis properties and individual phases evolution patterns. 
In sec.\ \ref{sec:MFR}, we present the multifractal analysis of the order
parameter fluctuations in two representative points of the hysteresis
loop. Final section \ref{sec:discuss} is devoted to a summary
and the discussion of the results.

\section{Structure of the 5-clique aggregates with different spectral
  dimension\label{sec:nets}}

\begin{figure*}[!htb]
\begin{tabular}{ccc} 
\resizebox{36pc}{!}{\includegraphics{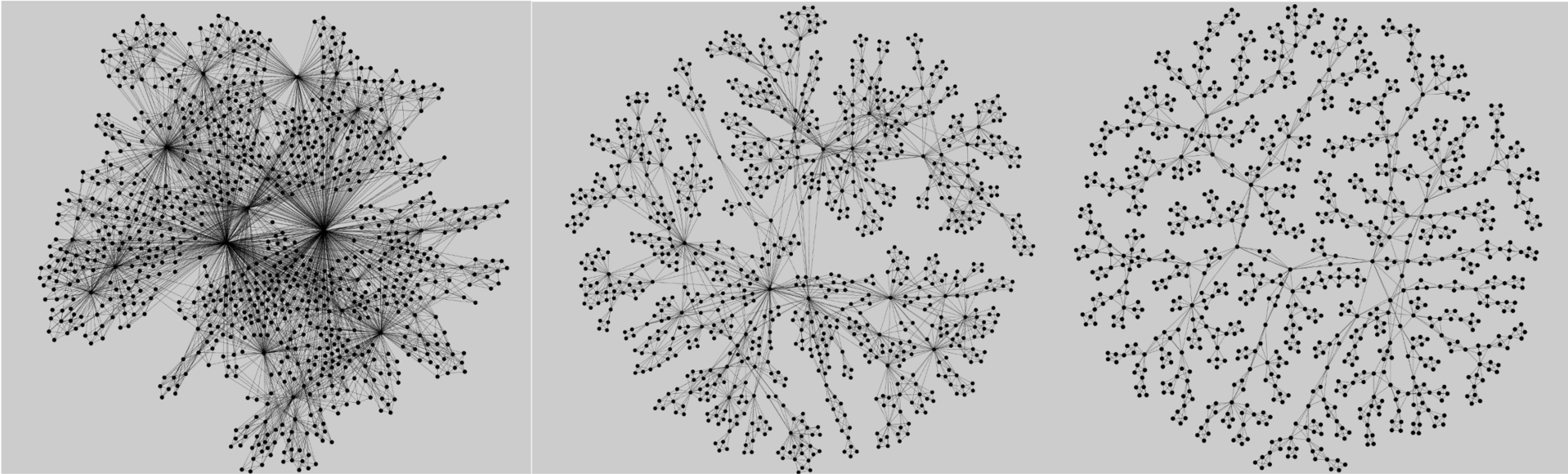}}\\
\end{tabular}
\caption{Networks of 5-cliques grown by the self-assembly rules
  described in \cite{SC_we_SciRep2018} observing the geometric compatibility for different
  chemical potential, left to right:  $\nu =5$ (compact), $\nu =0$ (mixed), and $\nu
  =-5$ (sparse structure). Adding 5-cliques is stopped when the number of nodes
  reaches $N\geq 1000$.}
\label{fig:nets3x}
\end{figure*}

\begin{figure*}[!htb]
\begin{tabular}{cccc} 
\resizebox{17pc}{!}{\includegraphics{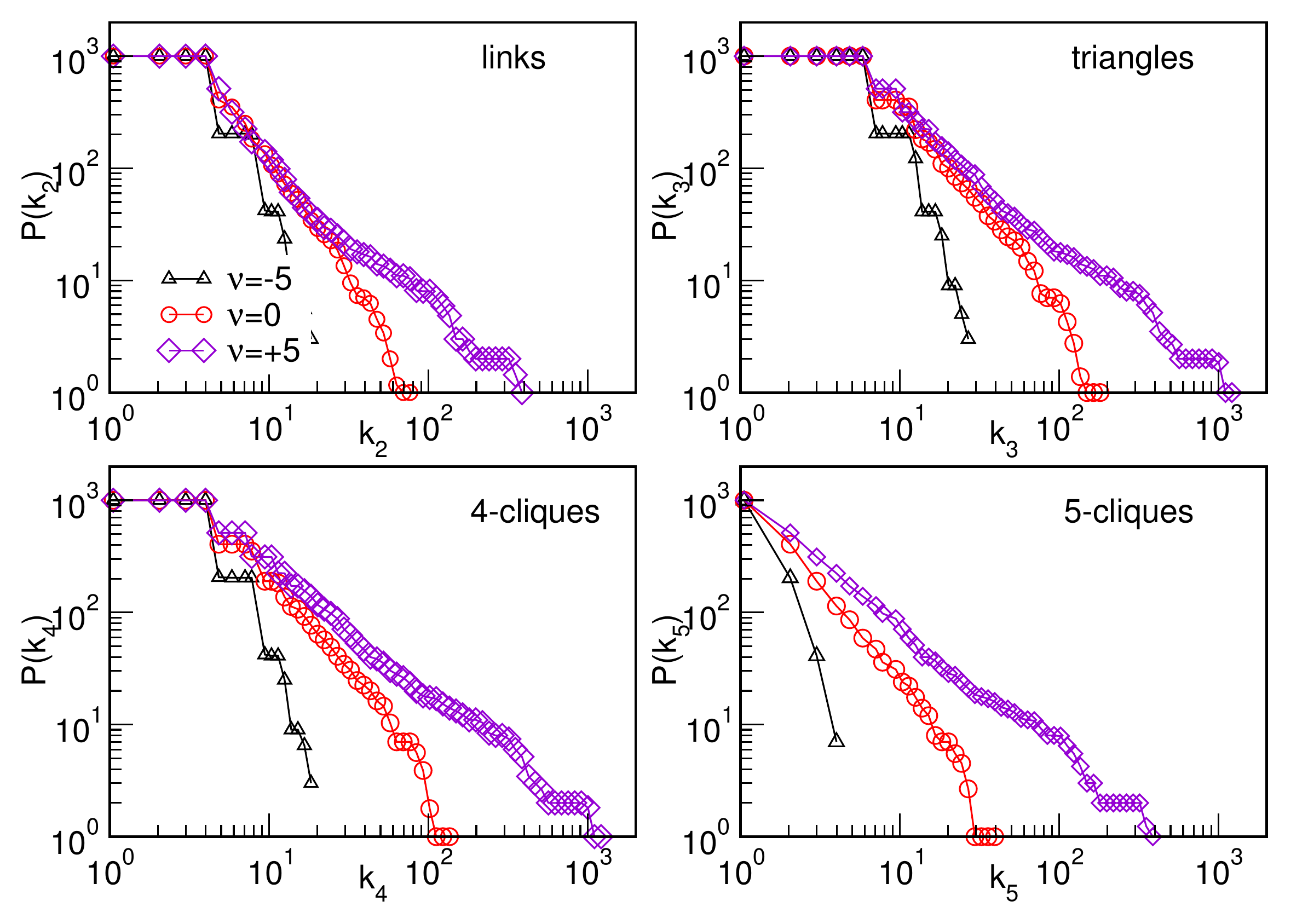}}&\resizebox{17pc}{!}{\includegraphics{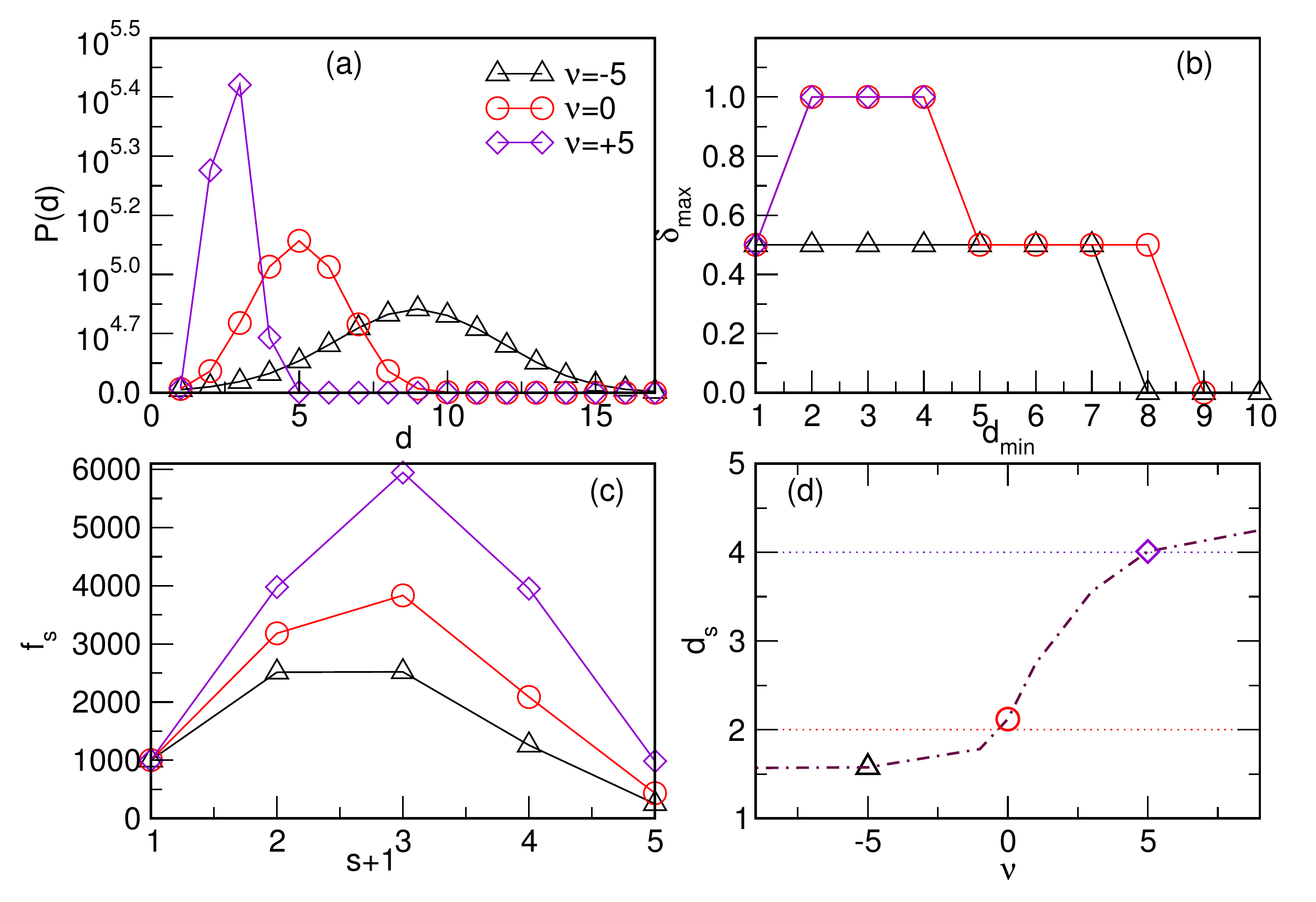}}\\
\end{tabular}
\caption{Left four panels: Cumulative distributions of the generalised
  degree: the number of edges, triangles, tetrahedra, and 5-cliques per node in three networks for different $\nu$, as indicated in the unique legend. Right four panels: (a) The distribution of the shortest-path distances between node pairs, $P(d)$ vd the distance $d$, (b) the hyperbolicity parameter $\delta_{max}$ vs the smallest distance $d_{min}$ of the nodes 4-tuples on these three networks, (c) the number of simplexes $f_s$ of different order $q$ in them, and (d) the network's spectral dimension $d_s$ for different $\nu$. The same legend applies to all three panels. The data in panel (d) are from the
reference \cite{SC_we-PREspectra2019}.}
\label{fig:netsDistrs}
\end{figure*}
To grow the 4-dimensional simplicial complexes for our study, we use 
the generative model   introduced in
\cite{SC_we_SciRep2018,we-ClNets-applet} for a cooperative
self-assembly of pre-formatted groups of nanoparticles
\cite{AT_cooperativeSA2021,AT_Materials_Jap2016}.  As explained in the Introduction, we
fix the size of the building blocks as 5-cliques; starting from a
single 5-clique, at each growth step, a new clique is attached to the
growing structure such that it shares one of its geometrical faces with a clique
which is already present in the structure. In the present case, the
possible faces are sub-cliques of the size $s=$1,2,3, and 4,
respectively, a
single node, a link with two adjacent nodes, a triangle with three
connected nodes, or a tetrahedron consisting of four nodes. What face would be shared is determined by the geometric compatibility factor
and the chemical affinity parameter $\nu$; see
Ref. \cite{SC_we_SciRep2018} for a detailed description, and \cite{SC_we_PRE2020} for an extended model with defect
cliques. Specifically, the
probability of sharing a face of the size $s$ is given by 
\begin{equation}
P(s_{max},s;t)= \frac{c_s(t)e^{-\nu (s_{max}-s)}}{\sum _{s=1}^{s_{max}-1}c_s(t)e^{-\nu (s_{max}-s)}} \ 
\label{eq-pattach}
\end{equation}
where $c_{s}(t)$ stands for the number of geometrically compatible
locations on the entire structure at the moment $t$ where  
docking a simplex of the size $s$ can be done. Note that, in the
present case, we have $s_{max}=5$ fixed. The geometric factor is
weighted by the chemical affinity $\nu$ 
\cite{SC_we_SciRep2018} towards new $s_{max}-s$ that must
be added to the structure 
after the face with $s$ nodes is shared with a previous clique. Hence,
 for a large $\nu>0$, sharing a maximal sub-clique (a tetrahedron) is favoured; the emergent structure is compact.
Whereas, when 
$\nu<0$, the probability of adding more nodes 
 is increasing. Thus,  for a large negative $\nu$, the cliques preferably share a
single node (minimal face), resulting in a sparse structure. Without chemical affinity factor, $\nu=0$, sharing of faces of any
size $s=1,\cdots 4$ can occur, subject to the geometric compatibility
factors alone. The resulting structures that we use here are for $\nu=-5$, $\nu=0$, and
$\nu=+5$, shown in Fig.\ \ref{fig:nets3x}.\\
The structure of these simplicial complexes is characterised by
several quantities, cf.\ Fig. \ref{fig:netsDistrs}, which are relevant to
the present study. We determine the generalised-degree distributions $P(k_\mu)$, where $k_\mu$, for $\mu=2,3,4,5$, indicates
the number of edges, triangles, tetrahedra and 5-cliques attached to a
node, see Fig.\ \ref{fig:netsDistrs} left four panels. The three
structures differ significantly in all of these
distributions. In particular, the nodes in the compact simplicial complex (at
$\nu=+5$) share a large number of simplexes of all orders, which leads
to a distorted power-law distribution at high $k_\mu$; meanwhile, the
distributions of the 
structure at $\nu=0$ with the same number of nodes are  nearly power-law with a (finite-size)
cut-off). On the other hand, the sparse system, grown with
$\nu=-5$ exhibits a fast decaying exponential distribution for all
simplex sizes.

The other measures are compatible with these features, shown in Fig.\ \ref{fig:netsDistrs}a-d.
Specifically, the number of simplexes-and-faces of
different sizes that are present in each simplicial structure, $f_s$, shown
in Fig.\ \ref{fig:netsDistrs}c, indicates that the compact simplicial
complex at $\nu=+5$ possesses the largest number of triangles, and
gradually the number of cliques of other sizes, compared to the
structure for $\nu=0$ and $\nu=-5$.  
The underlying network (1-skeleton of these simplicial complexes) 
exhibits some other properties that strongly vary with $\nu$. In
particular,  these are the distributions of the
shortest-path distances on the underlying graph, shown in Fig.\
\ref{fig:netsDistrs}a, and the spectral dimension, which is determined
for these graphs in ref.\ \cite{SC_we-PREspectra2019}, shown in Fig.\
\ref{fig:netsDistrs}d.  The spectral dimension for these three
representative structures varies, in particular,  $d_s=1.57$, $d_s=2.11$ and $d_s=4.01$,
for $\nu=-5$, $\nu=0$, and $\nu=5$, respectively; they are indicated by different symbols on the line
$d_s(\nu)$, which is determined from the Laplacian eigenvalues
distribution in \cite{SC_we-PREspectra2019} for a range of such
structures. Similarly, the distributions on these networks, cf.\ Fig.\
\ref{fig:netsDistrs}a, show that small distances prevail in the compact 
structure for $\nu=5$, peaking at $d_{max}=3$, whereas the peak moves
towards larger distances, i.e., $d_{max}=5$ and $d_{max}=10$ for
$\nu=0$ and $\nu =-5$, respectively.
On the other hand, these graphs are 1-hyperbolic by construction; see the discussion in the original work \cite{SC_we_SciRep2018}.
Specifically, because the cliques (which are $\delta_0=0$-hyperbolic objects) always share their faces in these structures, the hyperbolicity of the emergent complex is
given \cite{HB-BermudoGromov2013,Hyperbolicity_cliqueDecomposition2017} by $\delta_0 +1$. In Fig.\ \ref{fig:netsDistrs}b demonstrates it by numerically computing the
Gromov hyperbolicity parameter $\delta_{max}$, which does not exceed
one considering $10^9$ different 4-tuples on these networks.

\section{Phase synchronisation on 4-dimensional simplicial
  complexes\label{sec:synchro}}
\subsection{Dynamical model and simulations\label{sec:simulations}}
We consider an ensemble of $N$ coupled Kuramoto oscillators \cite{Rodrigues2016_sync_review} associated
with  the nodes of a given simplicial complex. The  equation governing
the evolution of the phase angle $\theta_i$  of  $i^{th}$ oscillator is given  by
\cite{SC_synchro_wePRE2021}  
\begin{eqnarray}
 \dot{\theta_i}  =  \omega_i + \frac{K_1}{k^{(1)}_i} \sum_{j=1}^{N}A_{ij} \sin{(\theta_j - \theta_i)} +  \nonumber \\
 +  \frac{K_2}{2 k^{(2)}_i} \sum_{j=1}^{N} \sum_{l=1}^{N} B_{ijl}\sin{(\theta_j + \theta_l - 2\theta_i)} \\\nonumber 
 \label{eq:dyn}
 \end{eqnarray}
where $\omega$'s are the intrinsic frequencies of the phase
oscillators. The second and third terms in Eq.~(\ref{eq:dyn}) represent
1-simplex and 2-simplex interactions, respectively. Note that
 three-node interactions of the $i$-th oscillator are based on
 each 2-simplex (triangle) incident on node $i$, thus introducing a natural generalization of the pair-wise interaction term \cite{SC_Sfboccaletti2021}.
Here, $A_{ij}$ is an element of the 1-simplex adjacency matrix
$\bf{A}$, such that $A_{ij}=1$ if nodes $i, j$ are connected by a link
and 0 otherwise. In the second term, $B_{ijl}$ is an element of the
2-simplex adjacency tensor $\bf{B}$, such that $B_{ijl}=1$ if nodes $i,
j, l$ belong to a common 2-simplex (triangle) and 0
otherwise. Likewise, the normalisation factors  $k_i^{(1)}$ and
  $k_i^{(2)}$ indicate the number of links and triangles of the node
    $i$, respectively; cf.\ Fig.\ref{fig:netsDistrs} for the structure of 
    the actual simplicial complexes.
The well-known Kuramoto order parameter can effectively quantify the degree of synchronization of the network
\begin{equation}\label{eq:op}
r = \left<\left|\frac{1}{N}\sum_{j=1}^{N}e^{i\theta_j}\right| \right>,
\end{equation}
where the brackets $\left<.\right>$ indicate the time average.

In the simulations, for each network node $i= 1, 2, \dots, N$, where
we have $N= 1000$, 1002, and 1003 corresponding to $\nu=5$, 0, and -5 networks, respectively, the initial conditions for $\theta_i$ are chosen randomly in
the range $\theta_i \in [0, 2\pi]$. The numerical solution for the set
of equations~\ref{eq:dyn} is performed using a numerical integration
algorithm ODEINT  from Python SciPy library~\cite{Math_PythonLib_Nat2020}. For each
set of parameter values, the system is iterated for 50 000 steps, with the time step $dt$ = 0.01 and 
always considering the previous state of each dynamical variable, the
procedure known as tracking the attractor \cite{Synch_RoyGupte} is
used in most hysteresis studies. 
The order parameter in Eq.~(\ref{eq:op}) is calculated in the asymptotic range considering the last 20 000 iterations. Further, to study the hysteresis properties, we track the system's trajectory as the coupling parameter $K_1$ is first adiabatically increased in steps $dK_1=0.1$ in the appropriate
range from negative to positive values, typically $K_1\in[-2,+2]$,
constituting the forward sweep, and then decreased along the backward
branch. Meanwhile, the strength of the higher-order interaction $K_2$
kept fixed, and the
internal frequencies of the oscillators $\omega_i$ are drawn from a
given distribution, as explained below.

\subsection{Partial synchronisation at $K_1< 0$: Hysteresis loop
for uniform internal frequencies \label{sec:omega1}}
\begin{figure}[!htb]
\begin{tabular}{ccc} 
\resizebox{18pc}{!}{\includegraphics{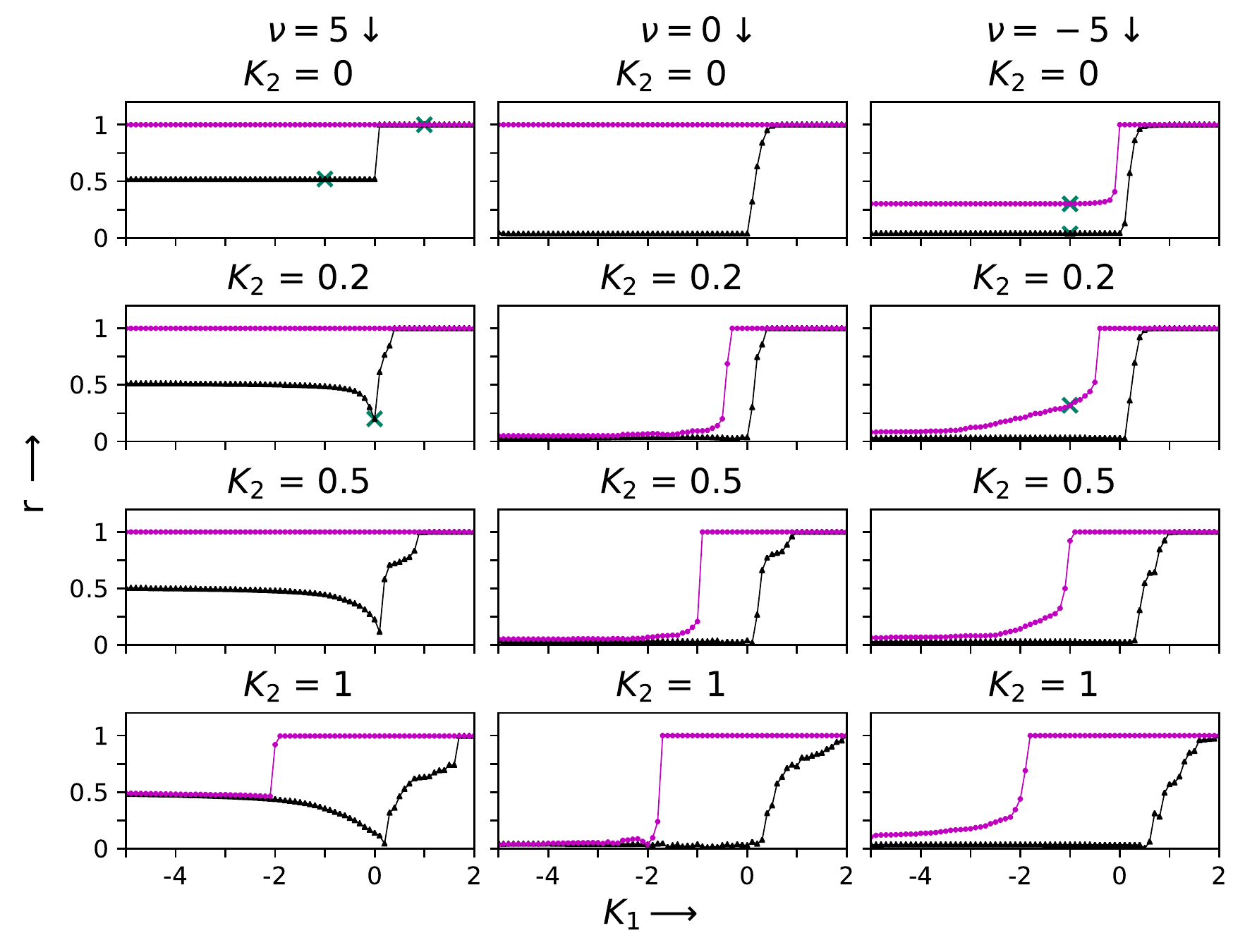}}\\
\end{tabular}
\caption{(Colour online) Hysteresis sweep of the order parameter $r$
  as a function of 1-simplex coupling strength $K_1$ at different
  2-simplex coupling strength $K_2$;   the three vertical columns
  (from left to right) correspond to three simplicial complexes in
  Fig.\ \ref{fig:nets3x} grown with the chemical affinity $\nu=$5, 0 \& -5, respectively. In each panel, the solid triangles
  (black) \& solid circles (magenta) refer  to forward \& backward
  sweeps, respectively. The intrinsic frequencies $\omega_i \eqsim
  1.0$ at all  nodes. 
The phase evolution patterns analysed
in \ref{sec:patterns} correspond to the points indicated by crosses.}
\label{fig:opOmega1}
\end{figure}

 In this section, we study hysteresis loop  for the interactions embedded
 in simplexes of different architectures described in sec.\
\ref{sec:nets}, when the internal
 frequencies of all oscillators are equal, i.e., $\omega\eqsim 1.0$ drawn from
 a $\delta-$function distribution. In this way, we expect that the
 impact of the structure and related interactions will be more pronounced.
As described above in \ref{sec:simulations}, for a given value of $K_2$ and
varying the pairwise coupling strength $K_1\in[-5,+2]$ in smal steps, the
order parameter is computed first along the forward branch, and then back.
The resulting hysteresis loops for the
three networks of Fig.\ \ref{fig:nets3x}, and fixing $K_2$  to several representatives values between $K_2=0.0$ and  $K_2=1.0$, are summarised in Fig.\ \ref{fig:opOmega1}.

As  Fig.\ \ref{fig:opOmega1} shows, even though the internal
frequencies of the oscillators are equal at all nodes, the shape of the hysteresis loop significantly depends on the structure of the underlying simplicial complexes. On the forward sweep,  by keeping $K_2=0$, we note the occurrence of partial synchronisation at 
negative $K_1$ values, where the order parameter reaches $r\approx 0.5$,
for the compact simplicial complex for $\nu=5$. 
In the other two networks ($\nu=0$ and $\nu=-5$),  
a much smaller but nonzero (within numerical error bars) value $r \approx 0.03$ is observed.   Further
increasing $K_1>0$, a smooth transition to a complete synchronisation with $r\approx 1$ occurs in all networks. \\
On the backward sweep, we note that the synchronised state persists,
and the loop does not close even at very large negative $K_1$ unless the higher-order coupling of a given strength $K_2>0$ is applied; moreover, the needed higher-order interaction correlates with the network's compactness. Specifically,
in the compact network ($\nu=5$), cf.\ lower left panel, 
the loop closes for $K_2=1.0$ via sudden desynchronisation, in
agreement with previous studies \cite{Arenas_NatureComm,HOC-Synch_Ginestra2020,SC_Sfboccaletti2021,SC_synchro_wePRE2021}. However,  for $\nu=-5$, an incomplete abrupt desynchronisation occurs even at $K_2=0.0$. Still, the loop closes gradually, reaching the forward branch for more negative $K_1$ values, cf.\ top right panel and the panels below it. In the intermediate case, $\nu=0$, a small $K_2=0.2$ suffices to induce an incomplete desynchronisation, and the loop gradually closes at  more
negative $K_1$ values. By increasing the triangle-based coupling
$K_2$, the apparent broadening of the loop superimposes these
features in each particular network. It is also accompanied by the slower reaching the full synchrony even at high positive pairwise
couplings. 
In the following, we will focus on the evolution of the phases of all oscillators at specific values $(K_1, K_2)$.

\begin{figure*}[!htb]
\begin{tabular}{ccc} 
\resizebox{32pc}{!}{\includegraphics{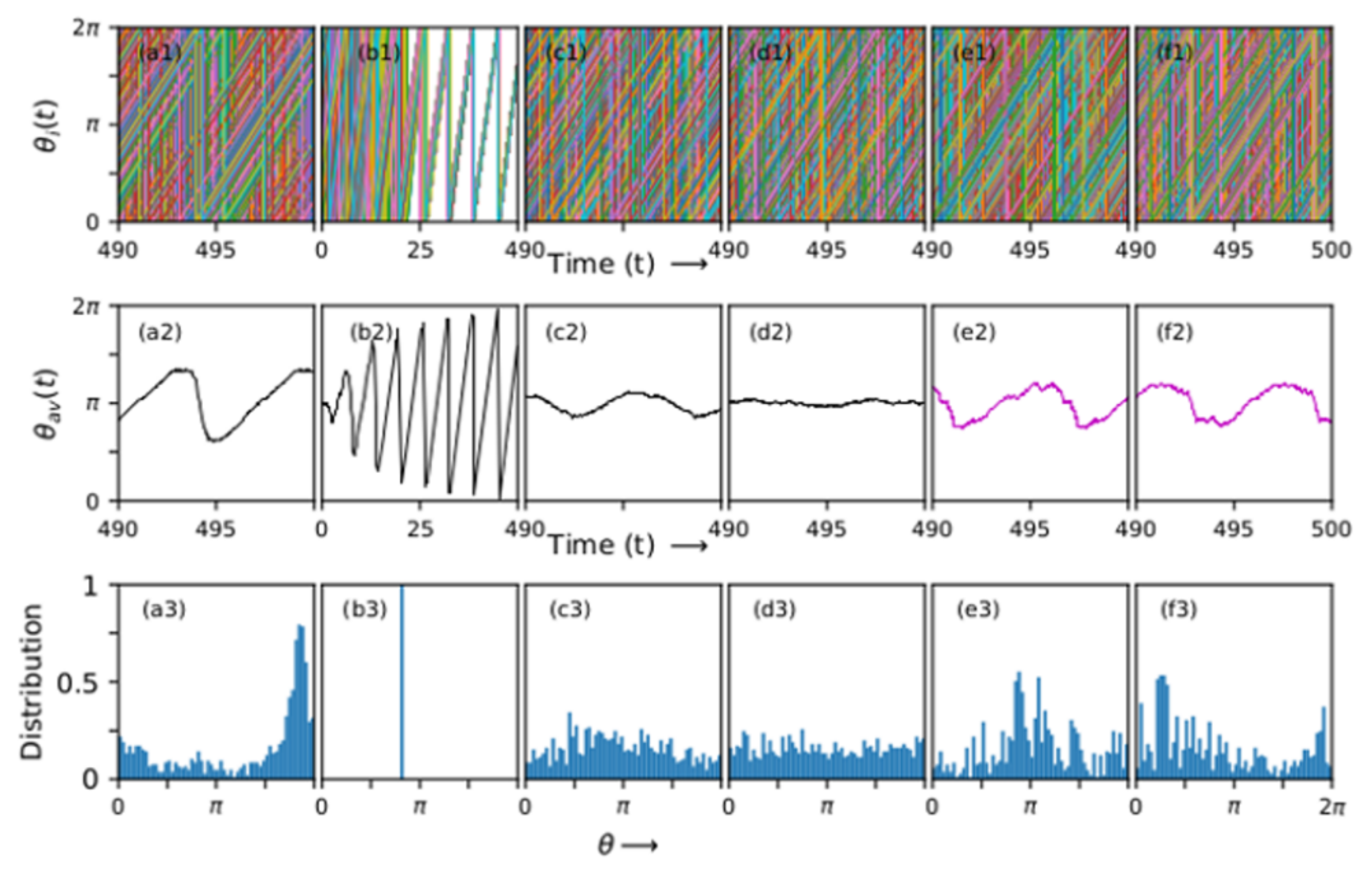}}\\
\end{tabular}
\caption{Colour online) Top row: Individual node's dynamics $\theta_i(t)$ vs
  time $t$ in the indicated interval (final 1000 time steps, except
  for the pattern (b1), which is shown for the initial 1000 time
  steps).  The parameter $(K_1, K_2)f,b$ on the forward  or backward branch are set as  $(-1,0)f$,  $(1,0)b$, and$(0,0.2)f$, for the network $\nu=5$, corresponding to patterns (a1), (b1), and (c1), respectively; meanwhile, the patterns (d1), (e1)
and (f1) are for $(-1,0)f$,
$(-1,0)b$, and$(-1,0.2)b$ for the sparse network with $\nu=-5$. Middle row: the network averaged phase $\theta_{av}(t)$ as a function of time in the same time interval as the corresponding pattern above it. Bottom row: Histogram of the node's phases $\theta_i$ taken at the end of that period corresponding to the panels in the same column above it.
}
\label{fig:patterns2x}
\end{figure*}

%%%%%
\subsection{The dynamics of individual nodes and patterns behind partial
  synchronisation \label{sec:patterns} }
The global order parameter $r$ discussed above quantifies the extent to which the oscillators are synchronized at a given set of parameter values. 
To get a deeper insight into  the synchronization and desynchronization processes and their dependence on network geometry and interactions, we study the global phase
angle $\theta_{av}(t)=\sum_i\theta_i(t)/N$ as a function of time, the evolution of phase angles of
each node, and the distribution of phase angles at a
particular time; cf.\ Fig.\ \ref{fig:patterns2x}.

 The phase trajectories of individual nodes are shown in the top row of
 Fig.~\ref{fig:patterns2x}; they are taken at different points marked by crosses on the hysteresis loops in Fig.~\ref{fig:opOmega1}. In particular, these points correspond to the values of the two interactions $(K_1, K_2)f,b$ on the forward $f$ or backward $b$ hysteresis branch are 
$(-1,0)f$,  $(-1,0)b$, and$(0,0.2)f$ for the compact network $\nu=5$, patterns (a1), (b1) and (c1), respectively. Similarly, the patterns on the
panels (d1), (e1) and (f1) correspond to the case $(-1,0)f$,
$(-1,0)b$, and $(-1,0.2)b$ for the sparse network with $\nu=-5$. 
The middle and bottom row below each pattern shows the corresponding
network's averaged phase $\theta_{av}(t)$ in the same time interval,
and the histogram of phases of individual nodes at the end of that
time interval. 

For the compact structure ($\nu=5$), we have that the order parameter $r\approx 0.5$ at the point  $(-1,0)f$; the corresponding pattern of phases, shown in the panel (a1) of Fig.\ \ref{fig:patterns2x}, indicates that
groups of roughly synchronised nodes are formed and evolve with the
same speed. The respective average phase fluctuates in
an extended range around $\pi$, as shown in panel (a2). In the
distribution of nodes' phases,  in (a3), broad peaks indicate
the formation of three groups of nodes with close but not fully
synchronised phases $\theta_i$.
The situation is much different at $(-1,0)b$ in the backward branch,
where the system remains fully synchronised, corresponding to a single sharp peak in (b3). The pattern of individual phases in
(b1) shows how such a synchronised state forms  in time when starting from a
random initial state. Consequently, the average phase in (b2) reaches the
full range of values on the unit circle. 
The panels (c1--c3) show how the order parameter appears at the point
$(0,0.2)f$ under the impact of weak triangle-based coupling alone. The pattern in (c1) and the corresponding average phase in (c2) show that
the network's compactness promotes ordering by forming small
groups, even though the leading pairwise interaction is absent.

The phase evolution patterns are different in the sparse network ($\nu=-5$), as shown in the panels (d1--f1).
At the point $(-1,0)f$, the order parameter value $r=0.03$
appears through many small groups of (roughly) synchronised
nodes, shown in (d1), corresponding to an almost even distribution of phases over the network nodes, cf. (d3), with tiny fluctuations of the average phase about $\pi$, in (d2). At the point  $(-1,0)b$ on the backward branch, the order parameter is dropped from $r=1$ to a finite value, which is compatible with the pronounced formation of groups visible in the panel (e1)  and in the distribution, cf.\ (e3). The corresponding average phase fluctuates in a larger interval, as shown in the panel (e2). A similar fluctuation range of the average phase is observed in the panel (f2), corresponding to the similar value of the order
parameter at the point $(-1,0.2)b$, cf.\ Fig.\
\ref{fig:opOmega1}. However, the presence of a weak higher-order
interaction, in this case, leads to a different grouping of nodes,
which is illustrated by the phase evolution pattern in the panel (f1) and
the phase histogram in (f3).

\subsection{Hysteresis properties in the transition from uniform
  to distributed internal frequencies\label{sec:gauss}}
\begin{figure}[!htb]
	\centering
        \resizebox{18pc}{!}{\includegraphics{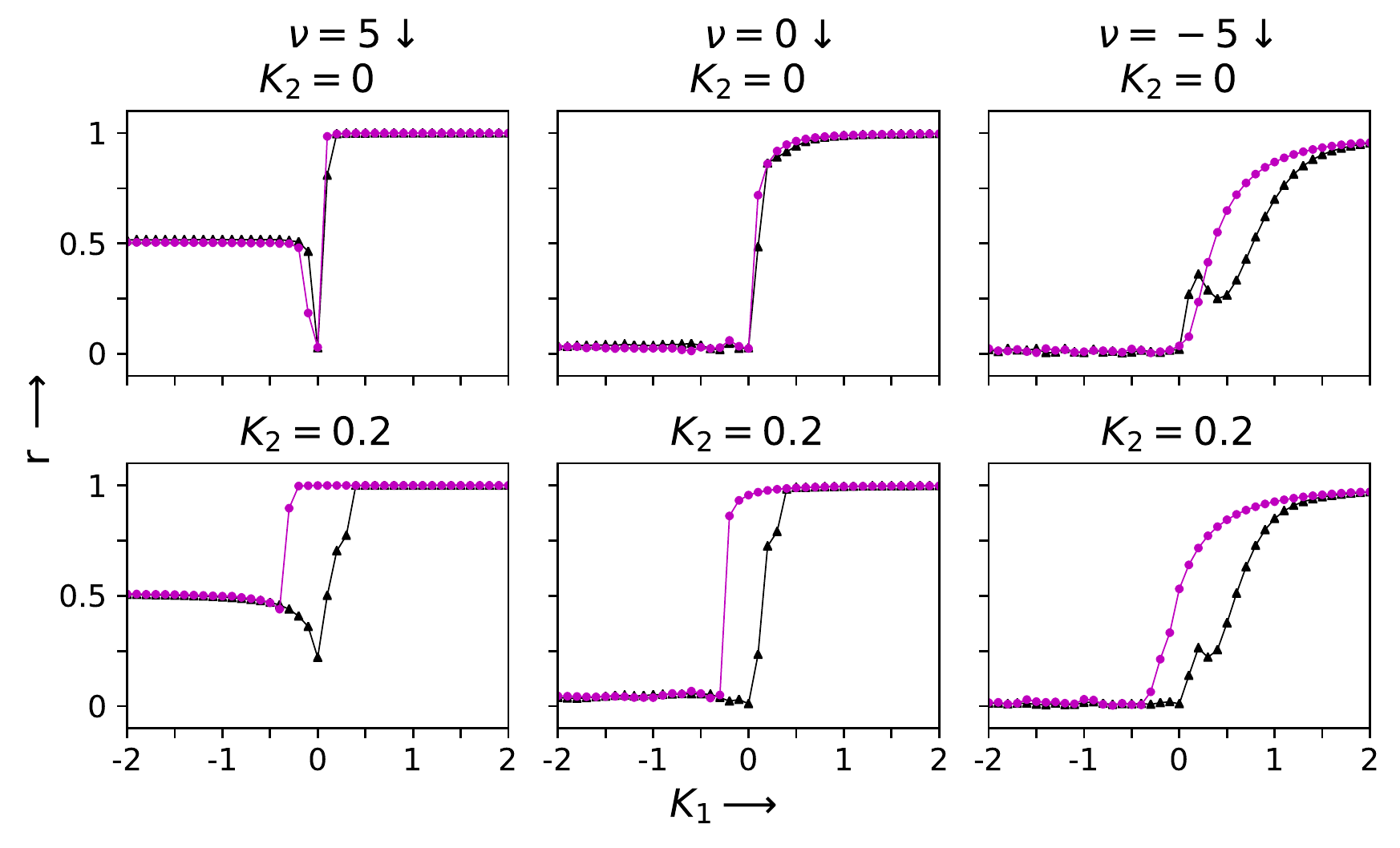}}\\
	\caption{(Colour online) Transition to synchronization \&
          hysteresis in networks with $\nu=$5, 0, -5 when, left to
          right columns, where $\omega$ is drawn from a Gaussian
          distribution width $\sigma=$ 0.01 around $\omega=1.0$;
          $K_2=0$ (top row), and  $K_2=0.2$ (bottom row).
The solid triangles (black) \& solid circles (magenta) represent the value of order parameter $r$ in forward \& backward sweeps, respectively.}
	\label{fig:HLomegas001}
\end{figure}

\begin{figure}[!htb]
	\centering
\resizebox{18pc}{!}{\includegraphics{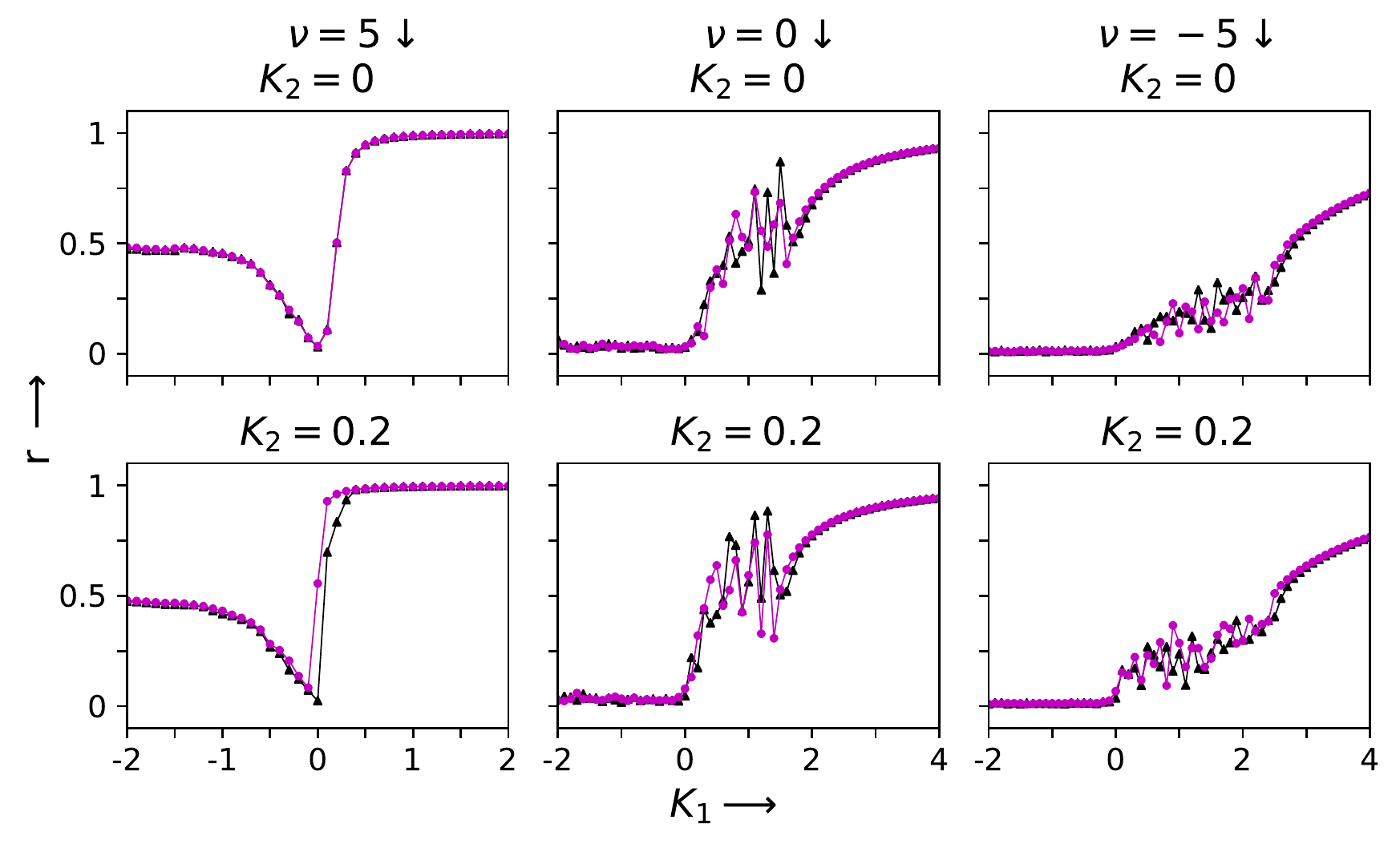}}\\
	\caption{(Colour online) Same as Fig.\ \ref{fig:HLomegas001} but with the
          $\omega$ drawn from a Gaussian distribution with the width
          $\sigma=0.1$. In the sparse network, we find that $r\to 1$ asymptotically
          at $K_1\gtrsim 10$ (not shown).   
Note the absence of  hysteresis for these
          values of higher-order interactions $K_2$.}
	\label{fig:HLomegaGauss01}
\end{figure}

To explore the impact of the distribution of internal oscillator frequencies,  they are drawn from a normal distribution of the width $\sigma$ centred about $\omega=1.0$. By increasing the width of the distribution $\sigma$, we show that new features of the hysteresis loop appear compared to the case of $\delta$-distribution in
Fig.\ \ref{fig:opOmega1}, and how these features depend on the
network structure. The results for the intermediate width, $\sigma
=0.01$,  and a broad distribution $\sigma
=0.1$, are shown in Fig.\ \ref{fig:HLomegas001} and Fig.\ \ref{fig:HLomegaGauss01}, respectively.

Both Fig.\ \ref{fig:HLomegas001} and Fig.\ \ref{fig:HLomegaGauss01} show that, in the presence of distributed internal frequencies, the partial synchronisation at $K_1<0$ persists in all networks with the respective value of the order parameter unchanged compared to the case of
$\delta-$distribution, cf.\ Fig.\ \ref{fig:opOmega1}. Moreover, the hysteresis loop is virtually absent unless the higher-order coupling
$K_2$ is switched on. Remarkably lower values of $K_2$ are needed to induce desynchronisation via an abrupt drop of the order parameter, i.e., in the compact network ($\nu=5$) compared with the case of homogeneous internal frequencies. We also observe several new features in the forward/backward sweeps. Particularly in the compact network, a drop of the order parameter at $K_1\leq 0$ occurs before it raises
again to reach complete synchrony at $K_1>0$; see the left columns
in Fig.\ \ref{fig:HLomegas001} and Fig.\ \ref{fig:HLomegaGauss01}.  While such a drop is absent in the case of $\delta-$ distribution, the area $K_1\lesssim 0$ where the order parameter is decreasing from the level $r=0.5$ to zero is
broadening with the increasing the frequency distribution width
$\sigma$. In the sparse and mixed networks, on the other hand, the
increasing spread of internal frequencies width $\sigma$ makes the 
complete synchronisation increasingly more difficult even at very
large $K_1>0$. For example, when $\sigma =0.1$, the pairwise coupling
$K_1\approx 10$ is required for the order parameter to approach
$r\lesssim 1$ (not shown). Instead of a steep increase in synchrony for
$K_1\geq 0$, we observe a characteristic instability where the (time-averaged) order parameter achieves different values when the interaction strength $K_1$ is changed by a small amount; this feature appears both at the forward as well as the backward branch and is practically unaffected by the weak higher-order interactions; see the middle and the right columns in Fig.\
\ref{fig:HLomegaGauss01}.  To a smaller extent, such instability is
seen in the sparse network already at a smaller distribution width,
$\sigma=0.01$, where it causes a kind of hysteresis at the positive
$K_1$ side, as shown in the top right panel of Fig.\ \ref{fig:HLomegas001}.
We note that this feature appears in the networks  where the building cliques share a single node, which is 100\% in the case of
sparse network, and also present in the mixed network, but entirely
absent in the compact network; cf.\ Fig.\ \ref{fig:nets3x}.
Understanding the mechanisms of how these instabilities appear is another challenging problem. In the next section, we analyse the nature of the order parameter fluctuations for the values of interactions in the range where the instability occurs in the sparse network.

\begin{figure*}[!htb]
\begin{tabular}{ccc} \resizebox{16pc}{!}{\includegraphics{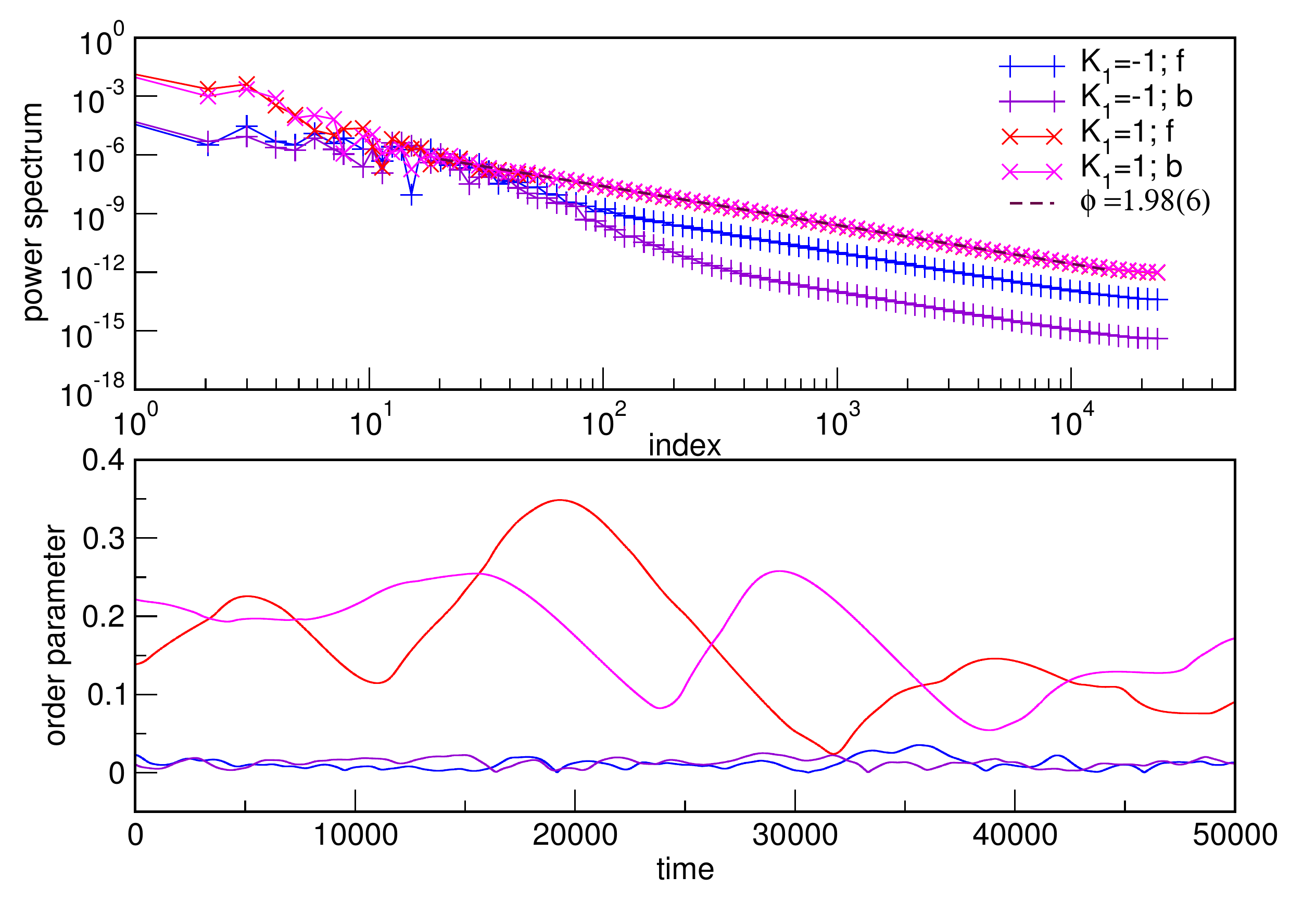}}&
\resizebox{24pc}{!}{\includegraphics{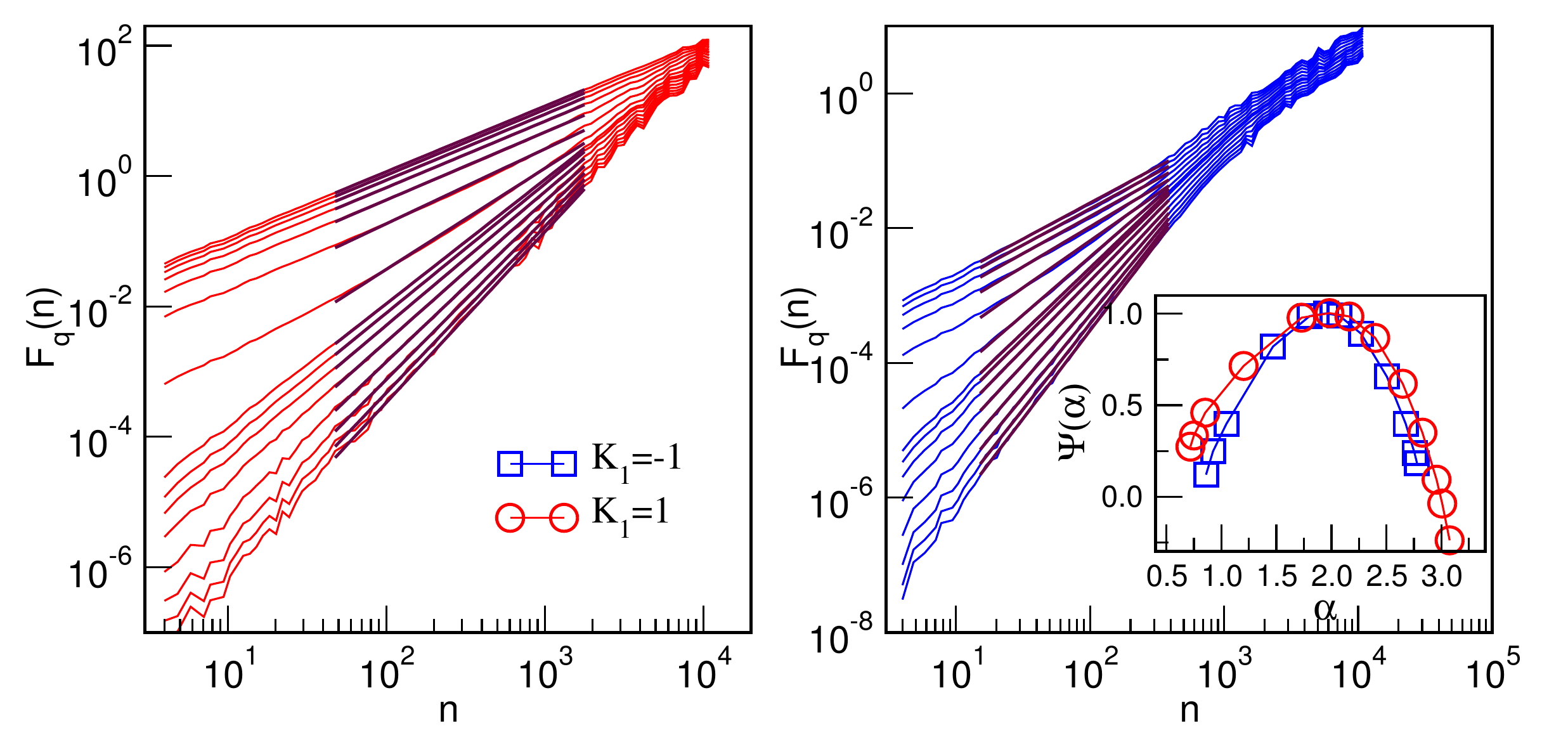}}\\
\end{tabular}
\caption{Left: In the network for $\nu =-5$, the order parameter vs
  time at two points corresponding to the partial synchronisation at $K_1=-1$ and $K_1=1$ on the
  forward (f) and backward (b) loop in the absence of higher-order
  coupling $K_2=0$. Middle and right: The fluctuation function
  $F_q(n)$ vs the time interval $n$ for the order-parameter in the
  forward loops at $K_1=-1$ (blue) and $K_1=1$ (red). Each line
  corresponds to different values of the amplification parameter
  $q\in [-3.5,3.5]$; the scaling area is indicated by the straight
  lines giving the generalised Hurst exponent $H_q$, which  leads to the corresponding singularity spectra $\Psi (\alpha)$ vs $\alpha$ given in the inset.}
\label{fig:RtPS_Fq2x}
\end{figure*}

\section{Multifractal fluctuations of the order parameter in partially synchronised
  states\label{sec:MFR}}

In the partially synchronised states in all simplicial complexes, the order parameter for $K_1<0$ has finite but different values; the time-averaged values are lower in the sparse networks than the compact ones. 
Here, we study temporal fluctuations of the order parameter for fixed pairwise interaction strength. Specifically, for the assembly at $\nu=-5$
and a broad Gaussian distribution  of the internal frequencies, partial synchronisation occurs at
$K_1<0$ but also for $K_1>0$,  reaching a full synchrony asymptotically
at very high $K_1$; cf.\ Fig.\
\ref{fig:HLomegaGauss01}; we consider two representative points, $K_1=-1$ and $K_1=+1$, in the absence of higher-order couplings. 
The respective time variations of the order parameter, shown in Fig.\
\ref{fig:RtPS_Fq2x} left panels, both for 
forward and backward branches of the hysteresis loop exhibit cyclical
fluctuations around different average values for $K_1<0$
compared to $K_1>0$. They lead to the exponent $\phi\sim 2$
compatible with the short-range correlations for an extended portion of
the power spectrum at large frequencies; cf.\ the top left panel in Fig.\
\ref{fig:RtPS_Fq2x}.   In the following, we show
that these cycles are modulated, attaining higher harmonics, which are
captured by the multifractal analysis.

For the analysis of the order parameter  fluctuations $r(t)$, we use
detrended multifractal analysis
of time series\ \cite{MFRA-uspekhi2007,DMFRA2002,we_BHN_MFR2016}.
Hence, the profile  $Y(i)
=\sum_{k=1}^i\ (r(k)-\langle  r\rangle)$ of the time series is divided in $N_s$ segments of the length $n$. 
Repeating the procedure starting from the end of the time
series$t=T_{max}$, we get  in total $2N_s=2Int(T_{max}/n)$ segments.
Then, at each segment $\mu=1,2\cdots N_s$, the local trend  $y_\mu(i)$
is determined, which allows computing the standard deviation around
it as  $F^2(\mu,n) =
\frac{1}{n}\sum_{i=1}^n\left[Y((\mu-1)n+i)-y_\mu(i)\right]^2$, and
similarly,  $F^2(\mu,n) = \frac{1}{n}\sum_{i=1}^n[Y(N-(\mu-N_s)n+i)-y_\mu(i)]^2$ for $\mu =N_s+1,\cdots 2N_s$.
The   fluctuation function $F_q(n)$ for the segment length $n$  is
then determined as 
\begin{equation}
F_q(n)=\left(\frac{1}{2N_s}\sum_{\mu=1}^{2N_s}\left[F^2(\mu,n)\right]^{q/2}\right)^{1/q} \sim n^{H_q}  \ ,
\label{eq:fluctq}
\end{equation}
for different positive and negative values of the amplification parameter $q$. 
The function is plotted against varied segment length
$n\in[2,int(T_{max}/4)]$. Its power-law sections on
the lines for different $q$ are fitted to find the
\textit{generalised Hurst exponent} $H_q$, defined on the
right-hand side of  the expression \ (\ref{eq:fluctq}). Notably, the case $q=2$
reduces  to the standard deviation function and corresponds to the
well-known Hurst exponent.

The spectrum $H_q$ is determined for a range of values of $q$ for
which the fluctuation function exhibits scale invariance; here, we use $q\in[-3.5,+3.5]$.
Once the spectrum $H_q$ is known, one can determine other
multifractality measures, in particular, $\tau_q=qH_q-1$, where the exponent  $\tau_q$, related to the
standard (box probability) measure \cite{DMFRA2002}. Then, using the Legendre
transform  $\Psi(\alpha)=q\alpha -\tau_q$, where $\alpha =d\tau/dq$
the  time
series \textit{singularity spectrum} can be determined. A nontrivial
singularity indicates different power-law singularities at different
data points $t$ of the time
series, according to $\vert \nabla
r(t,\epsilon)\vert _{\epsilon \to 0}\sim \epsilon ^{\alpha (t)}$ with
an exponent depending on the data point $t$
\cite{MFRA-uspekhi2007,DMFRA2002}.
Thus, the value  $\psi (\alpha)$ stands for a fractal dimension of the time series
points having  the same singularity exponent
$\alpha$. Note that for a monofractal, $H_q=H_2=const$, causing that the spectrum $\Psi(\alpha)$
reduces to a single point  $\alpha =H_2$, where $H_2$
is the standard Hurst exponent.

In Fig.\ \ref{fig:RtPS_Fq2x}, we show the results for the fluctuation
function $F_q(n)$ as a function of the time interval $n$ for the
order-parameter curves in the forward branch at $K_1=-1$ (blue) and $K_1=1$ (red lines). As these figures show, in both cases, the fluctuation function
$F_q(n)$ exhibits a scaling region for a broad range of time intervals
$n$. The fitted area of  $F_q(n)$ for different $q$ (indicated by
thick dark lines) gives the corresponding $H_q$ exponent defined in
eq.\ (\ref{eq:fluctq}). In both cases, the resulting broad spectra
$H_q$ are transformed onto the singularity spectra $\Psi(\alpha)$,
which are given in the inset in the right panel of Fig.\
\ref{fig:RtPS_Fq2x}. The parabolic distribution for both spectra is
asymmetrical, having a broad range of values with a maximum close to
$\alpha=2$ and somewhat
different curvature. Hence, the mechanisms behind the
occurrence of partial synchronisation, as discussed above, are
compatible with the multifractal temporal fluctuations of the order
parameter with broad singularity spectra.

\section{Discussion and Conclusions\label{sec:discuss}}

We have investigated the interplay of structure, interactions and
distribution of internal frequencies in phase synchronisation and desynchronisation processes on 4-dimensional simplicial complexes with different architecture composed of identical building blocks (5-cliques); cf.\ Fig.\ \ref{fig:nets3x}.
 Of the considered structures, 5-cliques are assembled with
rules of chemical affinity and geometric compatibility
\cite{SC_we_SciRep2018};
when chemical affinity allows the
addition of the maximum number of nodes,  a
minimal face (a node) is shared and a sparse structure appears; oppositely, 
sharing the largest face (4-clique) leads to a
compact assembly. For vanishing chemical affinity, a mixed
structure emerges where 5-cliques can share any of their faces by geometric compatibility.
The underlying graphs of these simplicial complexes are
1-hyperbolic and have different spectral dimensions \cite{SC_we-PREspectra2019},   as
shown in Fig.\ \ref{fig:netsDistrs}d.

Our results suggest that these simplicial architectures enable
geometric frustration effects and 
diverse collective dynamical phenomena. The shape of
the hysteresis loop in the presence of higher-order interactions, as
well as the collective fluctuations and the influence of the internal
frequency distribution on the synchronisation
processes on these simplicial complexes can be related to the 
size of shared faces by neighbouring 5-cliques. More precisely,
\begin{itemize}
\item Partial synchronisation $r<1$ occurs at negative pairwise
  coupling with a small non-zero value of the order parameter $r\sim 0.03$
  found when the least shared face matches one node ($s=1$);
  however, $r\sim 0.5$ when the least common face contains $s=4$
  nodes. 
Within numerical accuracy, these values of the global order parameter are insensitive to the internal frequency distribution and strength of triangle-based interactions.

\item Multiple interactions embedded in triangles change the hysteresis loop and,  at a strength that differs for each simplicial complex, induce a sudden drop of the order parameter to the corresponding partial synchronisation level on the negative branch of the pairwise interaction. 
Moreover, they prolong reaching the full synchronisation
with positive pairwise interactions in all simplicial complexes, the effect especially pronounced in the sparse architecture with distributed internal frequencies. Also, in the case of homogeneous frequencies, this structure allows incomplete abrupt desynchronisation even without higher-order coupling.

\item The evolution patterns of nodes, analysed at selected points on the hysteresis where partial synchronisation occurs, reveal co-evolving groups with different phases, which leads to quasi-oscillatory fluctuations of the order parameter. These fluctuations have multifractal features with broad singularity spectra associated with the simplicial structure and the interaction strength.
\end{itemize}
These findings shed new light on the nature of phase synchronisation in high-dimensional simplicial complexes of different architectures in
the interplay with coupling strengths and internal frequency distribution. While the transition to synchronisation induced by positive pairwise interactions was much investigated \cite{HOC_review2023boccaletti}, the
nature of partial order associated with negative interactions remains to be better understood.
In this regard, our simplicial complexes built of identical blocks as 4-dimensional cliques present a potential for studying the complexity of
synchronisation/desynchronisation processes in
greater detail beyond the measure of the spectral dimension. For example, studying the eigenvector localisation \cite{we-spectra2009} can reveal what
mesoscopic structures are involved and the role of node's correlations \cite{SC_synchro_wePRE2021,SC_we_HCsynchro2022} in the collective dynamics. Moreover, the influence of even/odd numbers of nodes in the shared faces and the geometry-embedded 4th- and 5th-order interactions remain open questions for future study.
The results presented here reveal mechanisms behind dynamic states with partial synchronisation, which are often desirable in complex functional systems, e.g., the brain, and ways to construct simplicial complexes of the same order that support full synchronisation when needed.

\section*{Acknowledgments}

B.T.  acknowledge the financial support from the Slovenian
Research Agency under the program P1-0044. S.S. acknowledges the financial support from IC\&SR, IITM through the project: SB20210838AMMHRD008291. N.G. thanks IC\&SR, IITM for the financial support through the project: SP20210777DRMHRDDIRIIT.

\bibliographystyle{unsrt}
%\bibliography{nets_synchro_.bib}

\end{document}